\documentclass[journal=jctcce,manuscript=article,layout=twocolumn]{achemso}

\usepackage[colorlinks,linkcolor=blue,citecolor=blue,urlcolor=black,bookmarks=false,hypertexnames=true]{hyperref} 

\usepackage[version=3]{mhchem}
\usepackage{amssymb}
\usepackage{amsmath}
\usepackage{color}
\usepackage{braket}
\usepackage{xspace}
\usepackage{cleveref}
\usepackage{graphicx}
\usepackage{subfigure}
\usepackage{threeparttable}
\usepackage{textcomp}
\usepackage{setspace}
\usepackage{caption}
\usepackage{comment}

\usepackage{array}
\newcolumntype{L}[1]{>{\raggedright\let\newline\\\arraybackslash\hspace{0pt}}m{#1}}
\newcolumntype{C}[1]{>{\centering\let\newline\\\arraybackslash\hspace{0pt}}m{#1}}
\newcolumntype{R}[1]{>{\raggedleft\let\newline\\\arraybackslash\hspace{0pt}}m{#1}}

\makeatletter
\let\l@addto@macro\relax
\makeatother
\usepackage[fontsize=11pt]{scrextend}

\let\oldmaketitle\maketitle
\let\maketitle\relax

\newcommand{\cm}{\ensuremath{\text{cm}^{-1}}\xspace}

\newcommand{\angstrom}{\mbox{\normalfont\AA}\xspace}

\DeclareUnicodeCharacter{2009}{\,}

\crefname{figure}{Figure}{Figures}
\crefname{table}{Table}{Tables}
\crefname{equation}{Eq.}{Eqs.}
\crefname{section}{Section}{Sections}
\crefname{subsection}{Section}{Sections}

\author{Rajat~Majumder$^\dag$}
\affiliation{$^\dag$Department of Chemistry and Biochemistry, The Ohio State University, Columbus, Ohio 43210, USA}

\author{Alexander~Yu.~Sokolov$^\dag$}
\email{sokolov.8@osu.edu}
\affiliation{$^\dag$Department of Chemistry and Biochemistry, The Ohio State University, Columbus, Ohio 43210, USA}

\begin{tocentry}
\includegraphics[width=1.0\textwidth]{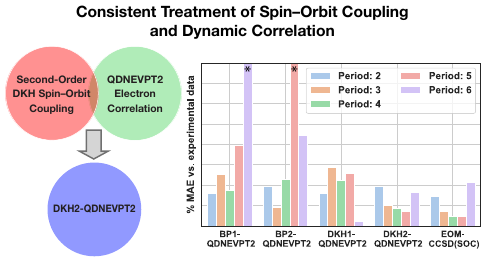}
\end{tocentry}

\title{{\color{blue}
Consistent Second-Order Treatment of Spin--Orbit Coupling and Dynamic Correlation in Quasidegenerate N-Electron Valence Perturbation Theory
}}

\begin{document}


\newcommand*{\abstractext}{
We present a formulation and implementation of second-order quasidegenerate N-electron valence perturbation theory (QDNEVPT2) that provides a balanced and accurate description of spin--orbit coupling and dynamic correlation effects in multiconfigurational electronic states.
In our approach, the energies and wavefunctions of electronic states are computed by treating electron repulsion and spin--orbit coupling operators as equal perturbations to the non-relativistic complete active-space wavefunctions and their contributions are incorporated fully up to the second order.
The spin--orbit effects are described using the Breit--Pauli (BP) or exact two-component Douglas--Kroll--Hess (DKH) Hamiltonians within spin--orbit mean-field approximation. 
The resulting second-order methods (BP2- and DKH2-QDNEVPT2) are capable of treating spin--orbit coupling effects in nearly degenerate electronic states by diagonalizing an effective Hamiltonian expanded in a compact non-relativistic basis.
For a variety of atoms and small molecules across the entire periodic table, we demonstrate that DKH2-QDNEVPT2 is competitive in accuracy with variational two-component relativistic theories.
BP2-QDNEVPT2 shows high accuracy for the second- and third-period elements, but its performance deteriorates for heavier atoms and molecules.
We also consider the first-order spin--orbit QDNEVPT2 approximations (BP1- and DKH1-QDNEVPT2), among which DKH1-QDNEVPT2 is reliable but less accurate than DKH2-QDNEVPT2.
Both DKH1- and DKH2-QDNEVPT2 hold promise as efficient and accurate electronic structure methods for treating electron correlation and spin--orbit coupling in a variety of applications.
\vspace{0.25cm}
}

\twocolumn[
\begin{@twocolumnfalse}
\oldmaketitle
\vspace{-0.75cm}
\begin{abstract}
\abstractext
\end{abstract}
\end{@twocolumnfalse}
]

\section{Introduction}
\label{sec:introduction}

Understanding and predicting many important properties of open-shell compounds require simultaneous description of spin--orbit coupling and electron correlation.
These properties include zero-field splittings, magnetic susceptibilities, intersystem crossing rates, phosphorescence lifetimes, core-level binding energies, and fine structure in X-ray or extreme ultraviolet light spectra.\cite{Kaszuba:1999p44274433,Woodruff:2013p51105148,McAdams:2017p216239,Lv:2021p214011,Lee:2010p9715,Kasper:2018p1998,Maganas:2019p104106,Carbone:2019p241,Stetina:2019p234103,Vidal:2020p8314} 
Rigorous treatment of open-shell electronic states can be achieved using the four-component relativistic theories based on the Dirac--Coulomb (DC) or  Dirac--Coulomb--Breit (DCB) Hamiltonians\cite{Jensen:1996p40834097,Liu:2010p1679,Saue:2011p3077,Fleig:2012p2,Pyykko:2012p45,Reiher:2004p10945,Kutzelnigg:2012p16} that introduce scalar and spin-dependent relativistic effects variationally in the mean-field wavefunction\cite{Gao:2004p66586666,Paquier:2018p174110,Yanai:2001p65266538,Sun:2021p33883402} and incorporate correlation by expanding the space of electronic and positronic configurations.\cite{Jensen:1996p40834097, Fleig:2009p1260712614,Liu:2021pe1536,Fleig:2006p104106,Abe:2008p157177,Abe:2006p234110,Saue:2011p3077, Shiozaki:2015p47334739,Reynolds:2019p15601571,Sun:2021p33883402,Hoyer:2023p44101}
Unfortunately, the four-component methods have much higher computational cost compared to their non-relativistic counterparts and their domain of applications remains rather limited.

Significant progress in achieving the accurate and balanced description of spin--orbit coupling and electron correlation in realistic chemical systems has been made by developing the two-component relativistic Hamiltonians,\cite{VanLenthe:1993p4597,Dyall:1997p96189626,Kutzelnigg:1997p203222,Kutzelnigg:2005p241102,Barysz:1997p225239,Reiher:2004p2037,Peng:2007p104106,Saue:2011p3077,Kutzelnigg:2012p16,Nakajima:2012p385,Li:2012p154114,Cheng:2014p164107}  such as the Breit--Pauli\cite{Breit:1932p616,Bearpark:1993p479502,Berning:2000p1823} (BP), zeroth-order regular approximation\cite{VanLenthe:1993p45974610,VanLenthe:1994p97839792, VanLenthe:1996p65056516} (ZORA), Douglas--Kroll--Hess\cite{Douglas:1974p89,Hess:1986p3742,Jansen:1989p6016} (DKH), the Barysz--Sadlej--Snijders\cite{Barysz:1997p225239,Sadlej:1998p1758} (BSS), and the exact two-component\cite{Kutzelnigg:2005p241102,Kutzelnigg:2006p22252240,Ilias:2007p064102,Li:2012p154114,Cheng:2014p164107} (X2C) Hamiltonians. 
By decoupling the physically relevant electronic states from the positronic degrees of freedom, the two-component methods achieve lower computational cost and can be more easily combined with the treatment of electron correlation effects compared to the four-component approaches.

The two-component relativistic theories can be broadly divided into two categories: (i) {\it variational} methods that incorporate relativistic effects in the self-consistent field (SCF) reference wavefunction\cite{Ganyushin:2013p104113,Mussard:2018p154,Jenkins:2019p2974,Lu:2022p29832992,Guo:2023p66686685,Wang:2023p848855} and (ii) {\it perturbative} approaches that introduce spin--orbit coupling as {\it a posteriori} correction together with dynamic correlation following a non-relativistic SCF calculation.\cite{finley:2001p042502,Roos:2004p29192927,Kleinschmidt:2006p124101,Cheng:2014p164107,Meitei:2020p3597} 
The variational two-component methods can accurately describe electron correlation and spin--orbit coupling in molecules with elements from the entire periodic system, but their computational cost remains considerably higher than that of non-relativistic theories.
On the other hand, the perturbative methods have much lower computational cost similar to that of non-relativistic methods, but may be unreliable for the compounds with heavier elements where the relativistic effects become particularly strong.
An alternative strategy is offered by the state-interaction approach based on quasidegenerate perturbation theory where the two-component relativistic Hamiltonian is diagonalized in the basis of selected non-relativistic electronic wavefunctions.\cite{Neese:1998p65686582,Malmqvist:2002p230,Neese:2005p034107,Sayfutyarova:2016p234301,Meitei:2020p3597} 
Although exact in the limit of full configuration interaction, this approach effectively treats spin--orbit coupling as the first-order perturbation and may require expressing the Hamiltonian in a large configuration space to obtain accurate results. 

In this work, we present a multireference quasidegenerate perturbation theory that incorporates spin--orbit coupling and dynamic correlation completely up to the second order, providing a cost-efficient and equal-footing treatment of these effects for electronic states with multiconfigurational electronic structures.
Our approach is based on the second-order quasidegenerate N-electron valence perturbation theory (QDNEVPT2),\cite{Angeli:2004p4043} which describes static and dynamic correlation in many electronic states simultaneously, free of intruder-state problems.\cite{Sokolov:2024parxiv}
Previous two-component implementations of QDNEVPT2 have been limited to the first-order treatment of spin--orbit coupling utilizing the BP relativistic Hamiltonian.\cite{Lang:2019p104104,Lang:2020p1025, Lang:2020p14109, Majumder:2023p546559}
Here, we employ the second-order Douglas--Kroll--Hess Hamiltonian (DKH2) in the exact two-component formulation\cite{Li:2014p054111} and incorporate all contributions from spin--orbit coupling and dynamic correlation effects up to the second order in perturbation expansion.
We demonstrate that this new approach performs consistently well for atoms and molecules across the entire periodic table and is significantly more accurate than the QDNEVPT2 methods with the first-order treatment of spin--orbit coupling. 

Our paper is organized as follows.
First, we will discuss the theory behind our new two-component QDNEVPT2 methods (\cref{sec:theory}). 
Next, we will provide a short overview of our implementation and discuss computational details (\cref{sec:comp_details}).
Following this, we  will benchmark the performance of our spin--orbit QDNEVPT2 methods  for the zero-field splitting in main group elements and diatomics (\cref{sec:soc_1}) and transition metal atoms (\cref{sec:soc_2}).
Finally, in \cref{sec:soc_3}, we will investigate the accuracy of QDNEVPT2 spin--orbit coupling treatment for challenging heavy element systems: uranium(V) ion (\ce{U^5+}), neptunyl dioxide (\ce{NpO2^2+}), and uranium dioxide (\ce{UO2^2+}). 
The summary of our findings and conclusions are provided in \cref{sec:conclusions}.

\section{Theory}
\label{sec:theory}

\subsection{Second-Order Quasidegenerate N-Electron Valence Perturbation Theory}
\label{sec:theory:nevpt}

Second-order quasidegenerate N-electron valence perturbation theory (QDNEVPT2)\cite{Angeli:2004p4043} is a multistate multireference approach that computes the dynamically correlated energies ($\textbf{E}$) of electronic states ($\textbf{Y}$)  by diagonalizing the matrix of effective Hamiltonian ($\boldsymbol{\cal{H}}_{\mathbf{eff}}$)
\begin{align}
	\label{eq:eff_H_eig_problem}
	\boldsymbol{\cal{H}}_{\mathbf{eff}}\, \textbf{Y} = \textbf{Y} \, \textbf{E}\ ,
\end{align}
expressed in the basis of complete active space self-consistent field (CASSCF) wavefunctions $\ket{\Psi_I^{(0)}}$.\cite{Hinze:1973p6424,Werner:1980p5794,Roos:1980p157,Werner:1985p5053,Siegbahn:1998p2384}

In the Hermitian QDNEVPT2 formulation,\cite{Kirtman:1981p798,Kirtman:2003p3890,Certain:2003p5977,Shavitt:2008p5711,Sharma:2016p034103,Lang:2019p104104,Lang:2020p14109} the matrix elements of $\boldsymbol{\cal{H}}_{\mathbf{eff}}$ have the form
\begin{align}
	\label{eq:eff_H_sym}
	\langle{\Psi_I^{(0)}}|\hat{\cal{H}}_{\mathrm{eff}}|{\Psi_J^{(0)}}\rangle 
	&= E^{(0)}_I \delta_{IJ}  + \langle{\Psi^{(0)}_{I}}|\hat{\cal{V}}|{\Psi^{(0)}_J}\rangle  \notag \\
	&+ \frac{1}{2} \langle{\Psi^{(0)}_{I}}|\hat{\cal{V}}|{\Psi^{(1)}_J}\rangle  \notag \\
	&+ \frac{1}{2} \langle{\Psi^{(1)}_{I}}|\hat{\cal{V}}|{\Psi^{(0)}_J}\rangle  \ ,
\end{align}
where $E^{(0)}_I$ is the CASSCF energy of $I$th electronic state, $\hat{\cal{V}}$ is the perturbation contribution to the electronic Hamiltonian $\hat{\cal{H}}$
\begin{align}
	\label{eq:perturbation_operator}
	\hat{\cal{V}} = \hat{\cal{H}} - \hat{\cal{H}}^{(0)}\ ,
\end{align}
and $\ket{\Psi^{(1)}_I}$ is the $I$th first-order correlated wavefunction
\begin{align}
	\label{eq:first_order_wfn}
	\ket{\Psi^{(1)}_I}
	&=  \frac{1}{E_I^{(0)} - \hat{\cal{H}}^{(0)}}  \hat{\cal{V}}  \ket{\Psi_I^{(0)}} \ .
\end{align}

The zeroth-order Hamiltonian $\hat{\cal{H}}^{(0)}$ appearing in \cref{eq:perturbation_operator,eq:first_order_wfn} is chosen to be the Dyall Hamiltonian\cite{Dyall:1998p4909,Sokolov:2024parxiv}
\begin{align}
	\label{eq:dyall_h}
	\hat{\cal{H}}^{(0)} = C + \sum_{i}\epsilon_{i}a^{\dagger}_{i}a_{i} + \sum_{a}\epsilon_{a}a^{\dagger}_{a}a_{a} +  \hat{\cal{H}}_{\mathrm{active}}
\end{align}
expressed in the basis of core (doubly occupied), active (frontier, partially occupied), and virtual (unoccupied) CASSCF spin-orbitals labeled with the ($i, j, k,l$), ($w, x, y, z$), and ($a, b, c, d$) indices, respectively.
In \cref{eq:dyall_h}, $\hat{\cal{H}}_{\mathrm{active}}$ contains all (one- and two-electron) active-space contributions to the full Hamiltonian $\hat{\cal{H}}$, making QDNEVPT2 resilient to the intruder-state problems. 
The orbital energies  $\epsilon_i$ and $\epsilon_a$  are computed as eigenvalues of the generalized Fock matrix.
Expressions for $\epsilon_i$, $\epsilon_a$, $\hat{\cal{H}}_{\mathrm{active}}$, and the constant term $C$ can be found elsewhere.\cite{Dyall:1998p4909,Angeli:2004p4043,Sokolov:2024parxiv}

To reduce the computational cost of calculating $\boldsymbol{\cal{H}}_{\mathbf{eff}}$, the first-order wavefunctions $\ket{\Psi^{(1)}_I}$ are approximated by introducing internal contraction
\begin{align}
	\label{eq:amp_nevpt2}
	\ket{\Psi^{(1)}_I} 
	\approx \sum_{\mu} t_{\mu I}^{(1)} \hat{\tau}_\mu \ket{\Psi_I^{(0)}}
	\equiv \sum_{\mu} t_{\mu I}^{(1)} \ket{\Phi_{\mu I}} \ ,
\end{align}	
which  projects $\ket{\Psi^{(1)}_I}$  onto the space of perturber functions $\ket{\Phi_{\mu I}}$ constructed by applying the two-electron excitation operators $\hat{\tau}_\mu$ to the zeroth-order states $\ket{\Psi_I^{(0)}}$.
As a result, the number of parameters $t_{\mu I}^{(1)}$ in the internally contracted  wavefunction $\ket{\Psi^{(1)}_I}$ grows much less steeply with increasing active space size as compared to the parameter space of uncontracted $\ket{\Psi^{(1)}_I}$, making the internally contracted QDNEVPT2 calculations more feasible for larger active spaces.
The amplitudes $t_{\mu I}^{(1)}$ are computed by solving the linear system of equations
\begin{align}
	\label{eq:amp_equation}
	\sum_{\nu} K_{\mu \nu I}  {t}^{(1)}_{\nu I} &= - \braket{\Phi_{\mu I}|\hat{\cal{V}}|\Psi^{(0)}_I} \ , \\
	\label{eq:amp_equation_K}
	K_{\mu \nu I} &= \braket{\Phi_{\mu I}|\hat{\cal{H}}^{(0)} - E_I^{(0)}|\Phi_{\nu I}}
\end{align}
and can be separated into eight excitation classes that are labeled by the number of electrons added to or removed from the active space upon excitation ($[0]$, $[\pm1]$, $[\pm2]$, $[0']$, and $[\pm1']$).\cite{Dyall:1998p4909,Angeli:2004p4043,Sokolov:2024parxiv} 
Two types of internal contraction have been implemented in QDNEVPT2: (i) strong contraction (sc) and (ii) full internal contraction (fic, also termed as partial contraction).\cite{Angeli:2001p10252,Angeli:2002p9138,Angeli:2006p054108} 
In this work, we employ the orbitally invariant and more accurate fic where more than one perturber function $\ket{\Phi_{\mu I}}$ is used for each excitation class.

QDNEVPT2 is a multistate formulation of state-specific N-electron valence perturbation theory (NEVPT2)\cite{Angeli:2001p10252,Angeli:2002p9138,Angeli:2006p054108} that accounts for the interaction between model states $\ket{\Psi_I^{(0)}}$ upon including dynamic electron correlation effects following the so-called diagonalize--perturb--diagonalize approach.\cite{Angeli:2004p4043, Sharma:2016p034103, Nishimoto:2020p137219}
The reference wavefunctions $\ket{\Psi_I^{(0)}}$ are obtained from the state-averaged CASSCF (SA-CASSCF) calculation where each model state is assigned a particular weight in the orbital optimization procedure. 
The dynamic correlation effects are represented by the perturbation operator $\hat{\cal{V}}$, which describes the electronic repulsion between electrons in non-active orbitals ($\hat{\cal{V}} = \hat{\cal{V}}_{\mathrm{ee}}$). 

In this work, we present a new formulation of QDNEVPT2 that treats the dynamic correlation {\it and} spin--orbit coupling effects on equal footing by incorporating the two-component spin--orbit Hamiltonian ($\hat{\cal{H}}_{\mathrm{SO}}$) into $\hat{\cal{V}}$ ($\hat{\cal{V}} = \hat{\cal{V}}_{\mathrm{ee}} + \hat{\cal{H}}_{\mathrm{SO}}$) and including all terms in the resulting perturbation expansion of the effective Hamiltonian up to the second order.  
Before we discuss this approach, we briefly introduce the three $\hat{\cal{H}}_{\mathrm{SO}}$ with different treatment of decoupling between electronic and positronic degrees of freedom that will be employed in our calculations.

\subsection{Two-Component Relativistic Hamiltonians}
\label{sec:theory:rel}

The starting point for our discussion of relativistic effects is the four-component Dirac equation for a particle with mass $m$\cite{Dyall:1995_book,Jensen:1996p40834097,Saue:2011p3077,Reiher:2014_book}
\begin{align}
	\label{eq:Dirac}
   \begin{pmatrix} 
      \mathbf{V_{ne}} & c \boldsymbol{\sigma} \cdot \mathbf{p} \\ 
      c\boldsymbol{\sigma}\cdot \mathbf{p} &  \mathbf{V_{ne}} - 2mc^{2}  
   \end{pmatrix}
   \begin{pmatrix}
   	\mathbf{\Psi^{L}} \\
   	\mathbf{\Psi^{S}}
   \end{pmatrix} 
   &= E
    \begin{pmatrix}
   	\mathbf{\Psi^{L}} \\
   	\mathbf{\Psi^{S}}
   \end{pmatrix} \ ,
\end{align}
where the Hamiltonian on the l.h.s.\@ depends on the electron-nuclear potential $\mathbf{V_{ne}}$, the particle's momentum $\mathbf{p}$, and a set of Pauli matrices $\boldsymbol{\sigma}$.
In \cref{eq:Dirac}, the eigenfunction of Dirac Hamiltonian is a four-component bispinor that is expressed in terms of its large ($\mathbf{\Psi^{L}}$) and small ($\mathbf{\Psi^{S}}$) two-component wavefunctions.

Introducing the nonretarded electron--electron interaction into the Dirac Hamiltonian gives rise to the Dirac--Coulomb--Breit (DCB) four-component Hamiltonian,\cite{Pyykko:2012p5464,Liu:2014p5989} which is expected to be sufficiently accurate for describing the chemical properties of many-electron systems.
However, obtaining the DCB eigenfunctions is significantly more computationally expensive than solving the non-relativistic Schr\"odinger equation due to the much larger size of many-body basis in the relativistic calculations. 

To reduce computational cost, several techniques for approximate decoupling of $\mathbf{\Psi^{L}}$ and $\mathbf{\Psi^{S}}$ have been developed, resulting in a variety of two-component relativistic Hamiltonians.\cite{VanLenthe:1993p4597,Dyall:1997p96189626,Kutzelnigg:1997p203222,Kutzelnigg:2005p241102,Barysz:1997p225239,Reiher:2004p2037,Peng:2007p104106,Saue:2011p3077,Kutzelnigg:2012p16,Nakajima:2012p385,Li:2012p154114,Cheng:2014p164107,Breit:1932p616,Bearpark:1993p479502,VanLenthe:1993p45974610,VanLenthe:1994p97839792, VanLenthe:1996p65056516,Barysz:1997p225239,Sadlej:1998p1758,Douglas:1974p89,Hess:1986p3742,Jansen:1989p6016}
We refer the readers to excellent publications on this topic\cite{Kutzelnigg:1997p203222,Kutzelnigg:2005p241102,Kutzelnigg:2012p16,Liu:2009p031104, Liu:2010p1679,Li:2012p154114,Li:2014p054111} and instead focus on the three two-component Hamiltonians that will be employed in our work: 1) Breit--Pauli Hamiltonian  (BP),\cite{Breit:1932p616,Bearpark:1993p479502,Berning:2000p1823} first-order Douglas--Kroll--Hess Hamiltonian (DKH1), and second-order DKH Hamiltonian (DKH2).\cite{Douglas:1974p89,Wolf:2002p9215,Reiher:2004p10945,Li:2012p154114} 

Each two-component Hamiltonian can be expressed as 
\begin{align}
	\label{eq:h_2c}
	\hat{\cal{H}}_{\mathrm{2c}} &= \hat{\cal{H}}_{\mathrm{SF}} + \hat{\cal{H}}_{\mathrm{SO}} \ ,
\end{align}
where $\hat{\cal{H}}_{\mathrm{SF}}$ is the spin-free contribution describing the scalar relativistic effects and $\hat{\cal{H}}_{\mathrm{SO}}$ is the spin-dependent component representing the spin--orbit and spin--spin coupling. 
The scalar relativistic effects are incorporated variationally in the reference SA-CASSCF calculation by including the one-electron $\hat{\cal{H}}_{\mathrm{SF}}$ as a contribution to the zeroth-order Hamiltonian $\hat{\cal{H}}^{(0)}$ (\cref{eq:perturbation_operator}).

In this work, in our definition of BP and DKH1 two-component Hamiltonians we choose $\hat{\cal{H}}_{\mathrm{SF}}$ to be the exact two-component spin-free one-electron (X2C-1e) Hamiltonian ($\hat{\cal{H}}_{\mathrm{SF}} = \hat{\cal{H}}_{\mathrm{SF}}^{\mathrm{X2C-1e}}$)\cite{Li:2012p154114} that provides a more accurate description of scalar relativistic effects than the spin-free BP and DKH1 Hamiltonians.
For the spin-free contribution to the DKH2 two-component Hamiltonian, $\hat{\cal{H}}_{\mathrm{SF}}^{\mathrm{X2C-1e}}$ is supplied with additional terms originating from the second-order transformation of one-electron spin-dependent operator (\cref{sec:theory:dk}) due to the picture change effect ($\hat{\cal{H}}_{\mathrm{SF}} = \hat{\cal{H}}_{\mathrm{SF}}^{\mathrm{X2C-1e}}+\hat{\cal{H}}_{\mathrm{SF}}^{\mathrm{DKH2}}$).
The working equations for $\hat{\cal{H}}_{\mathrm{SF}}^{\mathrm{X2C-1e}}$ and $\hat{\cal{H}}_{\mathrm{SF}}^{\mathrm{DKH2}}$ can be found in Ref.\@ \citenum{Li:2014p054111} and are not discussed here. 

Within the spin--orbit mean-field approximation (SOMF),\cite{Hess:1996p365, Berning:2000p1823} the spin-dependent Hamiltonian $\hat{\cal{H}}_{\mathrm{SO}}$ can be written in the general form:
\begin{align}
	\label{eq:h_2c_so_general}
	\hat{\cal{H}}_{\mathrm{SO}} &= i \frac{\alpha^2}{4} \sum_{\xi} \sum_{pq} F^{\xi}_{pq} \hat{D}^{\xi}_{pq} \ ,
\end{align}
where $\alpha = 1 / c$ is the fine-structure constant, the indices $(p,q,r,s)$ label all spatial molecular orbitals in the one-electron basis set, $\xi=x,y,z$ denotes Cartesian coordinates, and $\hat{D}^{\xi}_{pq}$ are the one-electron spin excitation operators
\begin{align}
	\hat{D}^{x}_{pq} &= {a}^{\dagger}_{p\alpha}{a}_{q\beta} + {a}^{\dagger}_{p\beta}{a}_{q\alpha} \ , \\
	\hat{D}^{y}_{pq} &= i ({a}^{\dagger}_{p\beta}{a}_{q\alpha} - {a}^{\dagger}_{p\alpha}{a}_{q\beta}) \ , \\
	\hat{D}^{z}_{pq} &= {a}^{\dagger}_{p\alpha}{a}_{q\alpha} - {a}^{\dagger}_{p\beta}{a}_{q\beta} 
\end{align}
with the labels $\alpha$ and $\beta$ denoting the spin-up and spin-down electrons, respectively.
The expressions for the matrix elements $F^{\xi}_{pq}$ of the BP, DKH1, and DKH2 two-component spin--orbit Hamiltonians are provided in \cref{sec:theory:bp,sec:theory:dk}.

\subsubsection{Breit--Pauli Hamiltonian}
\label{sec:theory:bp}

The Breit--Pauli (BP) spin--orbit Hamiltonian ($\hat{\cal{H}}^{\mathrm{BP}}_{\mathrm{SO}}$) is a two-component relativistic operator obtained from an analytic Foldy--Wouthuysen (FW) transformation\cite{Foldy:1950p29} of the four-component Dirac Hamiltonian with additional Coulomb and Gaunt two-electron terms.\cite{Breit:1932p616,Hess:1996p365,Berning:2000p1823} 
The matrix elements of $\hat{\cal{H}}^{\mathrm{BP}}_{\mathrm{SO}}$ within the SOMF approximation can be written as
\begin{align}
	\label{eq:bp_somf}
	F^{\mathrm{BP}, \xi}_{pq} = h^{\xi}_{pq} + \sum_{rs} P_{rs} \left(g^{\xi}_{pqrs} - \frac{3}{2}g^{\xi}_{sqpr} + \frac{3}{2}g^{\xi}_{spqr}\right) \ ,
\end{align}
where $P_{rs} = P_{r\alpha s\alpha} + P_{r\beta s\beta}$ is the spin-free one-particle density matrix of the reference SA-CASSCF wavefunction.
The one- and two-electron integrals  
\begin{align}
	\label{eq:intbp_1e_2e}
	h^{\xi}_{pq} &= -i \langle{\phi_p(1)}| \hat{h}_{\xi}(1) |{\phi_q(1)}\rangle \ , \\
	\label{eq:g_sso}
	g^{\xi}_{pqrs} &= -i \langle{\phi_p(1) \phi_r (2)}| \hat{g}_{\xi,\mathrm{sso}}(1,2) |{\phi_q (1)\phi_s (2)}\rangle
\end{align}
calculated in the spatial molecular orbital basis ($\phi_p$) represent the one-electron spin--orbit $\hat{h}_{\xi}(i)$ and the two-electron spin--same orbit $\hat{g}_{\xi,\mathrm{sso}}(i,j)$ operators
\begin{align}
	\label{eq:bp_h_1}
	\hat{h}_{\xi}(i) 
	&= \sum_{A}\frac{Z_{A}[\mathbf{r}_{iA}\times \mathbf{\hat{p}}(i)]_{\xi}}{r^{3}_{iA}} \ , \\
	\label{eq:bp_h_2}
	\hat{g}_{\xi,\mathrm{sso}}(i,j) &= -\frac{[\mathbf{r}_{ji}\times \mathbf{\hat{p}}(i)]_{\xi}}{r^{3}_{ij}} \ ,
\end{align}
where $Z_A$ is the charge of nucleus $A$, $\mathbf{r}_{ij}$ and $\mathbf{r}_{iA}$ are the coordinates of electron $i$ relative to electron $j$ and nucleus $A$, respectively, and $\mathbf{\hat{p}} (i)$ is the momentum operator for electron $i$.
The two-electron term of $F^{\mathrm{BP}, \xi}_{pq}$ in \cref{eq:bp_somf} also contains contributions from the spin--other orbit operator, which matrix elements can be fully expressed in terms of $g^{\xi}_{pqrs}$.\cite{Majumder:2023p546559}
The $g^{\xi}_{pqrs}$ integrals can be expressed more compactly in the standard Physicists' notation as:
\begin{align}
	\label{eq:g_sso_compact}
	g^{\xi}_{pqrs} 
	&= \sum_{o \pi} \epsilon^{\xi}_{o \pi} \langle{\phi_{p o}\phi_r}|{\phi_{q \pi}\phi_s}\rangle \ ,
\end{align}
where $\phi_{p o}= \frac{\mathrm{d} \phi_{p}}{\mathrm{d}  o}$ with respect to $o, \pi \in ({x,y,z})$ and $\epsilon^{\xi}_{o \pi}$ is the Levi-Civita symbol. 

The BP Hamiltonian is widely used to incorporate spin--orbit coupling effects in perturbative two-component electronic structure methods.
However, it is considered to be a low-$Z$ approximation that is valid when $Z^2 \alpha^2 \ll 1$, showing increasingly large errors for elements beyond the third row of periodic table.
A more accurate and systematically improvable description of relativistic effects is provided by the Douglas–Kroll--Hess (DKH) family of two-component Hamiltonians,\cite{Douglas:1974p89,Hess:1986p3742,Jansen:1989p6016,Li:2012p154114,Li:2014p054111} which we briefly review in \cref{sec:theory:dk}.
Due to their perturbative nature, the DKH Hamiltonians are well-suited for combinations with electronic structure method based on perturbation theory such as QDNEVPT2. 
For a more detailed discussion of DKH Hamiltonians, we refer to excellent Refs.\@ \citenum{Li:2012p154114} and \citenum{Li:2014p054111}.

\subsubsection{First- and Second-Order Douglas–Kroll--Hess Hamiltonians}
\label{sec:theory:dk}

The derivation of DKH two-component Hamiltonians starts by separating the four-component one-electron Dirac Hamiltonian into spin-free and spin-dependent contributions and block-diagonalizing the spin-free part in a kinetically balanced basis.\cite{Stanton:1984p19101918}
The spin-dependent terms are transformed to the block-eigenstate basis of spin-free Hamiltonian and are expanded perturbatively up to the order $n$, which defines the hierarchy of DKH$n$ two-component Hamiltonians ($\hat{\cal{H}}^{\mathrm{DKH}n}_{\mathrm{SO}}$). 
Here, we employ the DKH approach developed by Liu and co-workers where the block diagonalization of spin-free Hamiltonian is performed using the X2C-1e method,\cite{Li:2012p154114, Li:2014p054111} which provides a more accurate description of scalar relativistic terms ($\hat{\cal{H}}_{\mathrm{SF}} = \hat{\cal{H}}_{\mathrm{SF}}^{\mathrm{X2C-1e}}$) than that conventional DKH formulation.\cite{Douglas:1974p89,Hess:1986p3742,Jansen:1989p6016}
For $n > 1$, additional spin-free terms arise from the transformation of spin-dependent Hamiltonian due to the picture change effect, which are added to the X2C-1e spin-free Hamiltonian ($\hat{\cal{H}}_{\mathrm{SF}} = \hat{\cal{H}}_{\mathrm{SF}}^{\mathrm{X2C-1e}} + \hat{\cal{H}}^{\mathrm{DKH}n}_{\mathrm{SF}}$).

When represented in the form of \cref{eq:h_2c_so_general}, the matrix elements of DKH1 spin--orbit Hamiltonian can be expressed as:\cite{Li:2014p054111,Cao:2017p3713,Mussard:2018p154}
\begin{align}
\label{eq:dkh1_somf}
\mathbf{F}^{\mathrm{DKH}1,\xi} &= \mathbf{h}^{\mathrm{DKH}1,\xi} + \mathbf{g}^{\mathrm{DKH}1, \xi} \ , \\
\mathbf{h}^{\mathrm{DKH}1, \xi} &= \mathbf{R}^{\dagger}_{+}\mathbf{X}^{\dagger} \mathbf{h}^{\xi} \mathbf{X}\mathbf{R}_{+}  \ , \\
\mathbf{g}^{\mathrm{DKH}1,\xi} &= \mathbf{R}^{\dagger}_{+}(\mathbf{G}^{\mathrm{LL},\xi} + \mathbf{G}^{\mathrm{LS},\xi}\mathbf{X} + \mathbf{X}^{\dagger}\mathbf{G}^{\mathrm{SL},\xi} \notag \\
&+ \mathbf{X}^{\dagger}\mathbf{G}^{\mathrm{SS},\xi}\mathbf{X})\mathbf{R}_{+}  \ ,
\end{align}
where the matrix $ \mathbf{X}$ decouples $\mathbf{\Psi^{L}}$ and $\mathbf{\Psi^{S}}$ in \cref{eq:Dirac} using the X2C-1e approach. 
The $\mathbf{R_{+}}$ matrix accounts for the metric renormalization and is expressed as
\begin{align}
	\mathbf{R_{+}} &= \mathbf{S}_+^{-\frac{1}{2}}(\mathbf{S}_+^{-\frac{1}{2}}\tilde{\mathbf{S}}_{+}\mathbf{S}_+^{-\frac{1}{2}})^{-\frac{1}{2}}\mathbf{S}_+^{\frac{1}{2}}  \ ,\\ 
	\tilde{\mathbf{S}}_{+} &= \mathbf{S}_+ + \mathbf{X}^{\dagger} \mathbf{S}_{-}\mathbf{X}  \ , \\
	\mathbf{S}_{-} &= \frac{\alpha^2}{2} \mathbf{T}  \ ,
\end{align}
in terms of the non-relativistic overlap ($\mathbf{S}_+=\mathbf{S}$) and kinetic energy ($\mathbf{T}$) integrals.

The mean-field two-electron term $\mathbf{g}^{\mathrm{DKH}1,\xi}$ is defined in terms of the $\mathbf{G^{\mathrm{XY},\xi}}$ ($X,Y \in \{ \mathrm{L,S} \}$) matrices\cite{Li:2014p054111,Cao:2017p3713,Mussard:2018p154}
\begin{align}
\label{eq:GLL}
G^{\mathrm{LL},\xi}_{\rho \lambda} &= -\sum_{\mu \nu} 2K^{\xi}_{\mu \rho \nu \lambda} P^{\mathrm{SS}}_{\mu \nu}  \ , \\
\label{eq:GLS}
G^{\mathrm{LS},\xi}_{\rho \lambda} &= -\sum_{\mu \nu} (K^{\xi}_{\rho \mu \nu \lambda} + K^{\xi}_{\mu \rho \nu \lambda}) P^{\mathrm{LS}}_{\mu \nu } = - G^{\mathrm{SL},\xi}_{\lambda \rho}   \ , \\
\label{eq:GSS}
G^{\mathrm{SS},\xi}_{\rho \lambda} &= -\sum_{\mu \nu} 2(K^{\xi}_{\rho \lambda \nu \mu} + K^{\xi}_{\rho \lambda \mu \nu} - K^{\xi}_{\rho \mu \lambda \nu}) P^{\mathrm{LL}}_{\mu \nu}  \ ,
\end{align}
expressed in the atomic spin-orbital basis labeled with $\rho, \lambda, \nu, \mu$.
The two-electron spin--orbit integrals
\begin{align}
\label{Kint}
K^{\xi}_{\rho \lambda \nu \mu} &= \sum_{o \pi} \epsilon^{\xi}_{o \pi} \langle{\phi_{\rho o}\phi_{\nu \pi}}|{\phi_{\lambda} \phi_{\mu}}\rangle \end{align}
are related to $g^{\xi}_{\rho \lambda \nu \mu}$ in \cref{eq:g_sso_compact} via
\begin{align}
	g^{\xi}_{\rho \lambda \nu \mu} &= -(K^{\xi}_{\rho \lambda \nu \mu} + K^{\xi}_{\rho \lambda \mu \nu}) \ .
\end{align}  
The density matrices $\mathbf{P}^{\mathrm{SS}}$, $\mathbf{P}^{\mathrm{LS}}$, and $\mathbf{P}^{\mathrm{LL}}$ appearing in \cref{eq:GLL,eq:GLS,eq:GSS} are obtained from the spin-free SA-CASSCF density matrix  $\mathbf{P}$ (\cref{eq:bp_somf}):
\begin{align}
\mathbf{P}^{\mathrm{SS}} &= \mathbf{X}\mathbf{P}^{\mathrm{LL}}\mathbf{X}^{\dagger} \ , \\
\mathbf{P}^{\mathrm{LS}}& = \mathbf{P}^{\mathrm{LL}}\mathbf{X}^{\dagger} \ , \\
\mathbf{P}^{\mathrm{LL}} &= \frac{1}{2} \mathbf{R}_{+}\mathbf{{PR}}^{\dagger}_{+} \ .
\end{align}

In \cref{eq:GLL,eq:GLS,eq:GSS}, the $G^{\mathrm{LS},\xi}_{\rho \lambda}$ and $G^{\mathrm{SL},\xi}_{\lambda \rho}$ matrices describe the Coulomb-exchange interactions while $G^{\mathrm{LL},\xi}_{\rho \lambda}$ originates from the Gaunt-exchange terms.\cite{Cao:2017p3713}
The $G^{\mathrm{SS},\xi}_{\rho \lambda}$ matrix represents a mixture of direct Coulomb and Gaunt-exchange contributions.
Due to spin averaging, the direct Gaunt terms vanish.
The DKH1 Hamiltonian reduces to the BP Hamiltonian when $\mathbf{R}_{+} = \mathbf{1}$ and $\mathbf{X} = \mathbf{1}$.

Incorporating the second-order terms gives rise to the DKH2 spin--orbit Hamiltonian with matrix elements\cite{Li:2014p054111}
\begin{align}
\label{eq:dkh2_somf}
\mathbf{F}^{\mathrm{DKH}2,\xi} &= \mathbf{h}^{\mathrm{DKH}1,\xi} + \mathbf{h}^{\mathrm{DKH}2,\xi} + \mathbf{g}^{\mathrm{DKH}1, \xi} \ ,
\end{align}
where the second-order one-electron spin-dependent contribution $\mathbf{h}^{\mathrm{DKH}2,\xi}$ has the form:
\begin{align}
	\mathbf{h}^{\mathrm{DKH}2,\xi} &= \frac{4}{\alpha^{4}} (\vec{\mathbf{W}} \times \mathbf{T}^{-1}\vec{\mathbf{O}}^{\dagger} + \vec{\mathbf{O}}\times \mathbf{T}^{-1}\vec{\mathbf{W}}^{\dagger})^\xi
\end{align}
The components of vectors $\vec{\mathbf{W}}$ and $\vec{\mathbf{O}}$ are defined as:
\begin{align}
\mathbf{W}^{\xi} &= \frac{\alpha^2}{2}\mathbf{S}_{+}\mathbf{C}_{+}\mathbf{w}^{\xi}\mathbf{C}^{\dagger}_{-}\mathbf{T} \ , \\
w^{\xi}_{pq} &= -\frac{o^{\xi}_{pq}}{E_{-,q} - E_{+,p}} \ , \\
\mathbf{o}^{\xi} &= \mathbf{C}^{\dagger}_{+}\mathbf{O}^{\xi} \mathbf{C}_{-} \ , \\
\mathbf{O}^{\xi} &= \frac{\alpha^2}{4}\mathbf{R}^{\dagger}_{+}\mathbf{X}^{\dagger}\mathbf{h}^{\xi}\mathbf{R}_{-} \ ,
\end{align}
where $E_{+,p}$/$E_{-,p}$ and $\mathbf{C}_+$/$\mathbf{C}_-$ are the eigenvalues and eigenvectors obtained by solving the X2C-1e equations for the positive/negative energy states, respectively. 
The renormalization matrix $\mathbf{R}_{-}$ is given by:
\begin{align}
	\mathbf{R}_{-} &= \mathbf{S}_{-}^{-\frac{1}{2}}(\mathbf{S}_{-}^{-\frac{1}{2}}\tilde{\mathbf{S}}_{-}\mathbf{S}_{-}^{-\frac{1}{2}})^{-\frac{1}{2}}\mathbf{S}_{-}^{\frac{1}{2}} \ , \\
	\tilde{\mathbf{S}}_{-} &= \mathbf{S}_{-} + \tilde{\mathbf{X}}^{\dagger}\mathbf{S}_{+}\tilde{\mathbf{X}} \ , \\
	\tilde{\mathbf{X}} &= -\mathbf{S}_{+}^{-1}\mathbf{X}^\dag\mathbf{S}_{-} \ .
\end{align}
As for DKH1, the DKH2 contributions to the two-component spin--orbit Hamiltonian are computed using the decoupling matrix $\mathbf{X}$ obtained from the X2C-1e procedure.
The resulting sf-X2C-1e+so-DKH$n$ ($n$ = 1, 2) approach will be termed here as DKH$n$ for brevity. 

\subsection{Incorporating Spin--Orbit Coupling in QDNEVPT2}
\label{sec:theory:soqd}

To incorporate spin--orbit coupling in QDNEVPT2, we augment the perturbation operator $\hat{\cal{V}}$ with a two-component spin--orbit Hamiltonian ($\hat{\cal{V}} = \hat{\cal{V}}_{\mathrm{ee}} + \hat{\cal{H}}_{\mathrm{SO}}$). 
The resulting effective Hamiltonian expanded up to the second order in perturbation theory has the form:
\begin{align}
	\label{eq:eff_H_sym_SOMF2}
	\langle{\Psi_I^{(0)}}|&\hat{\cal{H}}_{\mathrm{eff,SO}}^{\mathrm{BP2/DKH2}}|{\Psi_J^{(0)}}\rangle 
	= E^{(0)}_I \delta_{IJ} \notag \\
	&+ \langle{\Psi^{(0)}_{I}}|\hat{\cal{V}}_{\mathrm{ee}} + {\hat{\cal{H}}}_{\mathrm{SO}}^{\mathrm{BP/DKH2}}|{\Psi^{(0)}_J}\rangle  \notag \\
	&+ \frac{1}{2} \langle{\Psi^{(0)}_{I}}|\hat{\cal{V}}_{\mathrm{ee}} + {\hat{\cal{H}}}_{\mathrm{SO}}^{\mathrm{BP/DKH2}} |{\tilde{\Psi}^{(1)}_J}\rangle  \notag \\
	&+  \frac{1}{2} \langle{\tilde{\Psi}^{(1)}_{I}}|\hat{\cal{V}}_{\mathrm{ee}} + {\hat{\cal{H}}}_{\mathrm{SO}}^{\mathrm{BP/DKH2}}|{\Psi^{(0)}_J}\rangle \ .
\end{align}
In this formulation that consistently treats dynamic correlation and spin--orbit coupling to second order, we choose $\hat{\cal{H}}_{\mathrm{SO}}^{\mathrm{BP/DKH2}}$ to be either the BP (\cref{eq:bp_somf}) or DKH2 (\cref{eq:dkh2_somf}) Hamiltonian in the form of \cref{eq:h_2c_so_general}, denoted as BP2-QDNEVPT2 or DKH2-QDNEVPT2, respectively.
Compared to conventional QDNEVPT2, the BP2/DKH2-QDNEVPT2 effective Hamiltonian contains new terms that depend on $\hat{\cal{H}}_{\mathrm{SO}}^{\mathrm{BP/DKH2}}$ and modified first-order wavefunctions
\begin{align}
	\ket{\tilde{\Psi}^{(1)}_I} &= \sum_{\mu} \tilde{t}^{(1)}_{\mu I} \ket{\Phi_{\mu I}}\ ,
\end{align}
which amplitudes are computed by solving the linear system of equations 
\begin{align}
	\label{eq:amp_equation_so}
	\sum_{\nu} K_{\mu \nu I}  \tilde{t}^{(1)}_{\nu I} &= - \braket{\Phi_{\mu I}|\hat{\cal{V}}_{\mathrm{ee}} + {\hat{\cal{H}}}_{\mathrm{SO}}^{\mathrm{BP/DKH2}}|\Psi^{(0)}_I}
\end{align}
with $K_{\mu \nu I}$ defined in \cref{eq:amp_equation_K}.
Due to mean-field spin--orbit approximation, the r.h.s.\@ of \cref{eq:amp_equation_so} has non-zero contributions from $\hat{\cal{H}}_{\mathrm{SO}}^{\mathrm{BP/DKH2}}$ only for the semi-internal $[0']$ and $[\pm1']$ excitations, making the corresponding $\tilde{t}^{(1)}_{\nu I}$  amplitudes complex-valued.
For the remaining excitation classes ($[0]$, $[\pm1]$, $[\pm2]$), \cref{eq:amp_equation_so} reduces to \cref{eq:amp_equation}, with $\tilde{t}^{(1)}_{\nu I} = {t}^{(1)}_{\nu I}$ where ${t}^{(1)}_{\nu I}$ are the conventional real-valued QDNEVPT2  amplitudes.
Since solving \cref{eq:amp_equation_so} involves inverting the matrix of shifted nonrelativistic Dyall Hamiltonian $K_{\mu \nu I}$ (\cref{eq:amp_equation_K}, also known as the Koopmans matrix), the BP2/DKH2-QDNEVPT2 methods are expected to be resilient to intruder-state problems, similar to the original QDNEVPT2 approach.

\begin{table*}[t]
	\caption{Two-component methods implemented in this work. 
		For each method, dynamic correlation (DC) and spin--orbit coupling (SO) are expanded to the order specified in the second and third column, respectively.
		Also indicated are the spin-free (SF) and SO Hamiltonians employed in each method.
		}
	\label{tab:methods}
	\setstretch{1}
	\small
	\centering
	\begin{threeparttable}
		\begin{tabular}{ccccc}
			\hline\hline
			Method 								& DC order & SO order	& SF Hamiltonian & SO Hamiltonian \\ 
			\hline
			BP1-QDNEVPT2 				& 2				&  1				& X2C-1e			 		& BP\\
			DKH1-QDNEVPT2			  & 2			  &  1				  &	X2C-1e						& DKH1 \\
			BP2-QDNEVPT2 				& 2				&  2 			   & X2C-1e	 			 		& BP \\
			DKH2-QDNEVPT2			  & 2			  &  2				 & X2C-1e + DKH2	   &  DKH2 \\
			\hline\hline \\
		\end{tabular}
	\end{threeparttable}
\end{table*}

In addition to BP2- and DKH2-QDNEVPT2, we also consider two approximations where the spin--orbit coupling is treated to first order in perturbation theory using either the BP or DKH1 Hamiltonians, abbreviated as BP1-QDNEVPT2 or DKH1-QDNEVPT2, respectively.
The corresponding effective Hamiltonian has the form: 
\begin{align}
	\label{eq:eff_H_sym_SOMF}
	\langle{\Psi_I^{(0)}}|&\hat{\cal{H}}_{\mathrm{eff,SO}}^{\mathrm{BP1/DKH1}}|{\Psi_J^{(0)}}\rangle 
	= E^{(0)}_I \delta_{IJ} \notag \\
	&+ \langle{\Psi^{(0)}_{I}}|\hat{\cal{V}}_{\mathrm{ee}} + {\hat{\cal{H}}}_{\mathrm{SO}}^{\mathrm{BP/DKH1}}|{\Psi^{(0)}_J}\rangle  \notag \\
	&+ \frac{1}{2} \langle{\Psi^{(0)}_{I}}|\hat{\cal{V}}_{\mathrm{ee}} |{{\Psi}^{(1)}_J}\rangle  \notag \\
	&+  \frac{1}{2} \langle{{\Psi}^{(1)}_{I}}|\hat{\cal{V}}_{\mathrm{ee}}|{\Psi^{(0)}_J}\rangle \ ,
\end{align}
where $\ket{\Psi^{(1)}_I}$ is the conventional QDNEVPT2 first-order wavefunction with real-valued amplitudes determined by solving \cref{eq:amp_equation}.
We note that the BP1-QDNEVPT2 method has been studied in detail in Ref.\@ \citenum{Majumder:2023p546559}, while the DKH1-QDNEVPT2 implementation is reported for the first time.
A summary of methods implemented in this work is provided in \cref{tab:methods}.

\section{Implementation and Computational Details}
\label{sec:comp_details}

The two-component relativistic methods outlined in \cref{tab:methods} were implemented in the development version of \textsc{Prism}.\cite{Sokolov:2023prism}
Our implementation utilizes full internal contraction, preserves the degeneracy of states with the same total angular momentum, and avoids the calculation of four-particle reduced density matrices using the techniques developed in Ref.\@ \citenum{Majumder:2023p546559}.
All integrals and the SA-CASSCF reference wavefunctions were computed using the \textsc{Pyscf} package.\cite{Sun:2020p024109}
In addition to \textsc{Pyscf}, \textsc{Prism} was interfaced with \textsc{Socutils},\cite{Wang:2022socutils} which provided the matrix elements of DKH1 Hamiltonian for the DKH1-QDNEVPT2 calculations. 
The DKH2 Hamiltonian matrix elements used in DKH2-QDNEVPT2 were implemented in a local version of \textsc{Socutils}. 

We benchmarked the performance of spin--orbit QDNEVPT2 methods for a variety of atomic and molecular systems.
All electrons were  correlated in all calculations (i.e., no frozen core approximation was invoked).
First, in \cref{sec:soc_1}, we assess their accuracy for calculating zero-field splitting in  main group elements and diatomics against the reference data from experiments and theory. 
For this study, all calculations were performed using the uncontracted ANO-RCC and ANO-RCC-VTZP basis sets.\cite{Roos:2004p28512858}
Other computational parameters (geometries, active spaces, number of states averaged in SA-CASSCF) are provided in the Supplementary Material.

Next, in \cref{sec:soc_2}, we use the spin--orbit QDNEVPT2 methods to calculate the ground- or excited-state zero-field splittings in transition metal atoms, namely: \ce{Sc}, \ce{Y}, \ce{La}, \ce{Ag}, and \ce{Au}. 
For all of these atoms, the all-electron X2C-TZVPall-2c basis set was used.\cite{Pollak:2017p3696}
The calculations of \ce{Sc}, \ce{Y} and \ce{La} in their ${}^2D$  ground states were performed with 3 electrons in 9 active orbitals (3e, 9o), which included the $ns$, $np$, and $(n-1)d$ shells with $n$ = 4, 5, and 6, respectively.
For  \ce{Ag} and \ce{Au}, we computed the excited ${}^2D$ zero-field splitting utilizing the (11e, 6o) active space corresponding to the $ns$ and $(n-1)d$ orbitals with $n$ = 5 and 6, respectively. 
Additional details of these calculations can be found in the Supplementary Material.

Finally, in \cref{sec:soc_3}, we test the performance of our two-component QDNEVPT2 methods for three chemical systems with strong relativistic effects: \ce{U^5+}, \ce{NpO2^2+}, and \ce{UO2^+}.
The calculations of \ce{U^5+} in its ${}^2F$ ground electronic term utilized the SARC-DKH2 basis set\cite{Rolfes:2020p18421849} and (1e, 7o) active space, which incorporated the $5f$ orbitals. 
For \ce{NpO2^2+}, the uncontracted ANO-RCC-VTZP and cc-pVTZ basis sets\cite{Dunning:1989p1007} were used for the \ce{Np} and \ce{O} atoms, respectively.
In the case of \ce{UO2^+}, the contracted ANO-RCC-VTZP basis set was employed for all atoms. 
Calculations of both molecules utilized the (7e, 10o) active space, as shown in the Supplementary Material. 
The \ce{NpO2^2+} and \ce{UO2^+} structures have linear geometries with the Np--O bond distance of 1.70 \angstrom and the U--O bond distance of 1.802 \angstrom. 

\section{Results and Discussion}
\label{sec:results}

\subsection{Main Group Elements and Diatomics}
\label{sec:soc_1}

\begin{table*}[t]
	\caption{
		Spin--orbit zero-field splitting (\cm) in the ${}^2P$ ground term of atoms and ${}^2\Pi$ ground term of diatomics computed using the spin--orbit QDNEVPT2 methods. 
		Results are compared to the reference data from the SO-EOM-CCSD method with relaxed amplitudes\cite{Cheng:2018p044108} and experiments.\cite{Moore:1949_book, Radziemski:1969p424443, LucKoenig:1975p199219, Martin:1979p817864, Huber:1982p298, Fink:1989p1928, Gilles:1991p47234724, Ram2000:p915, Miller:2001p312318, Drouin:2001p128138, Kramida:2007p544557,Shirai:2007p509615,Deverall:1953p297299}
		All methods employed the uncontracted ANO-RCC basis set.
	}
	\label{tab:small_mol}
	\setstretch{1}
	\footnotesize
	\centering
	\begin{threeparttable}
		\begin{tabular}{cccccccc}
			\hline\hline
			System			& BP1-		& BP2-		& DKH1-		& DKH2-			& SO-			& Experiment	\\ 
			& QDNEVPT2	& QDNEVPT2	& QDNEVPT2	& QDNEPVT2		& EOM-CCSD\cite{Cheng:2018p044108}		&			\\			
			\hline
			\ce{B} 			& 15.0		& 14.5		& 15.0			& 14.5		& 13.7			& 15.3\cite{Kramida:2007p544557}\\	
			\ce{Al}			& 107.6		& 109.9		& 106.8		& 109.4		& 107.5			& 112\cite{Martin:1979p817864} 	\\
			\ce{Ga}			& 887.4		&  867.9		& 840.4		& 818.8		& 797.6			& 826\cite{Shirai:2007p509615}	\\
			\ce{In} 			& 2560.8		& 2859.2 		& 2205.2		& 2219.0		& 2103.6			& 2213\cite{Deverall:1953p297299}	\\
			\ce{Tl} 			& 12475.8		& 8655.5 		& 7745.1 		& 8113.3		& 6794.1			& 7793\cite{Moore:1949_book}	\\\\
			\ce{F} 			& 401.5		& 405.7 		& 400.5		& 405.0		& 396.8			& 404\cite{Moore:1949_book}	\\
			\ce{Cl}			& 789.7		& 867.8 		& 779.5 		& 858.6		& 872.8			& 882\cite{Radziemski:1969p424443}	\\
			\ce{Br}			& 3574.4		& 3926.0 		& 3329.4	 	& 3625.0		& 3555.4			& 3685\cite{Moore:1949_book}	\\
			\ce{I} 			& 8149.9		& 10343.7 		& 6824.7		& 7581.0		& 7288.8			& 7603\cite{LucKoenig:1975p199219}	\\\\
			\ce{OH}	    	     	& 152.5      	& 123.4 		& 152.3		&  123.2         	& 136.3			& 139\cite{Huber:1982p298}	\\
			\ce{SH}			& 375.6		& 381.7 		& 371.4 		& 378.2		& 373.8			& 377\cite{Huber:1982p298}	\\
			\ce{SeH}			& 1836.7		& 1930.1		& 1719.5		& 1793.2		& 1716.8			& 1763\cite{Ram2000:p915}	\\
			\ce{TeH}			& 4293.5		& 5238.1		& 3637.4		& 3956.5		& 3751.7			& 3816\cite{Fink:1989p1928}	\\\\
			\ce{FO} 			& 180.0		& 189.5		& 179.5		& 189.2		& 193.6			& 197\cite{Miller:2001p312318}	\\
			\ce{ClO}			& 299.7		& 326.6 		& 297.0 		& 324.4		& 318.7			& 322\cite{Drouin:2001p49}	\\
			\ce{BrO}			& 961.9		& 1085.4		& 903.5	 	& 1012.0		& 984.2			& 975\cite{Drouin:2001p128138}	\\
			\ce{IO}			& 2303.8		& 2924.2 		& 1959.7 		& 2237.5		& 2143.6			& 2091\cite{Gilles:1991p47234724}	\\
			\hline\hline
		\end{tabular}
	\end{threeparttable}
\end{table*}

\begin{figure*}[t]
	\subfigure[]{
		\includegraphics[width=0.48\textwidth]{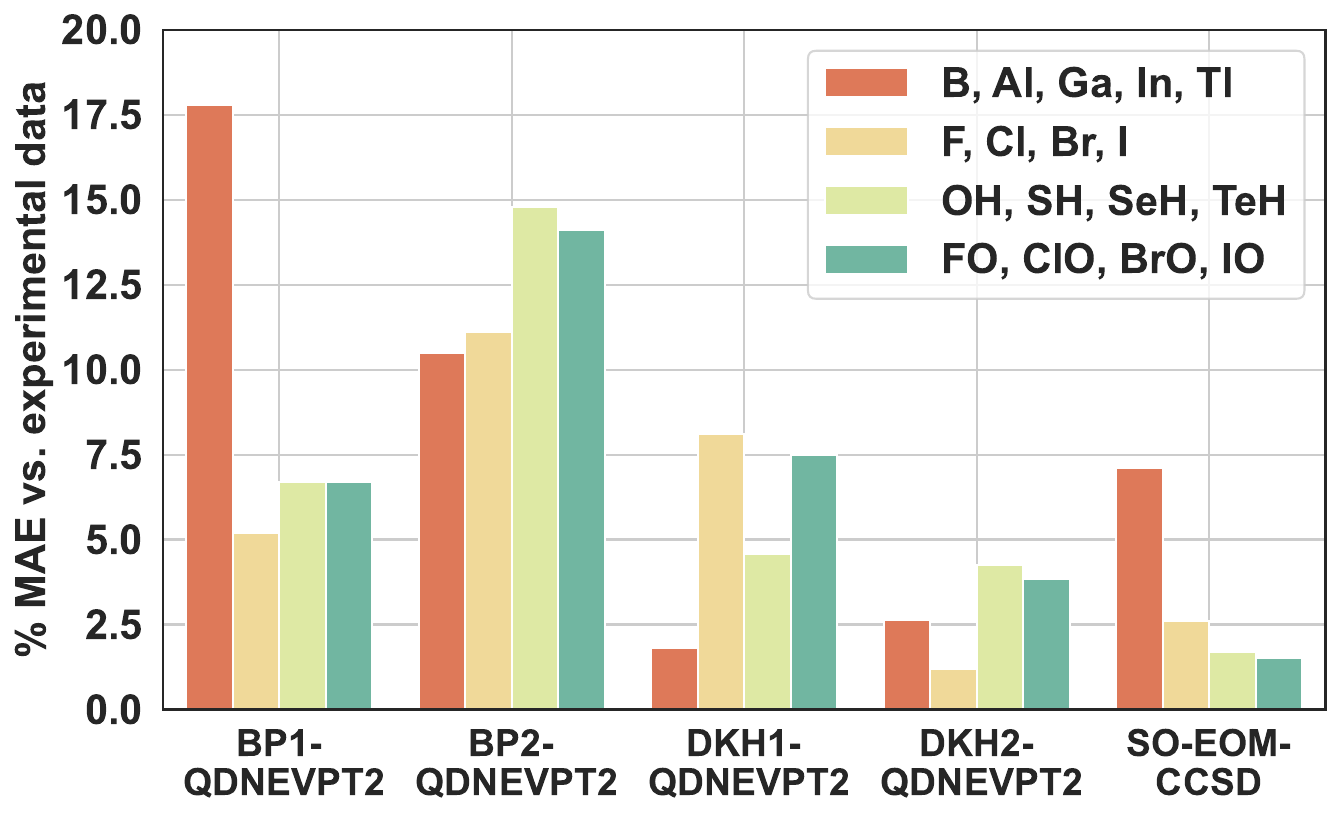}\label{figa:pblock_lan}}  
	\subfigure[]{
		\includegraphics[width=0.48\textwidth]{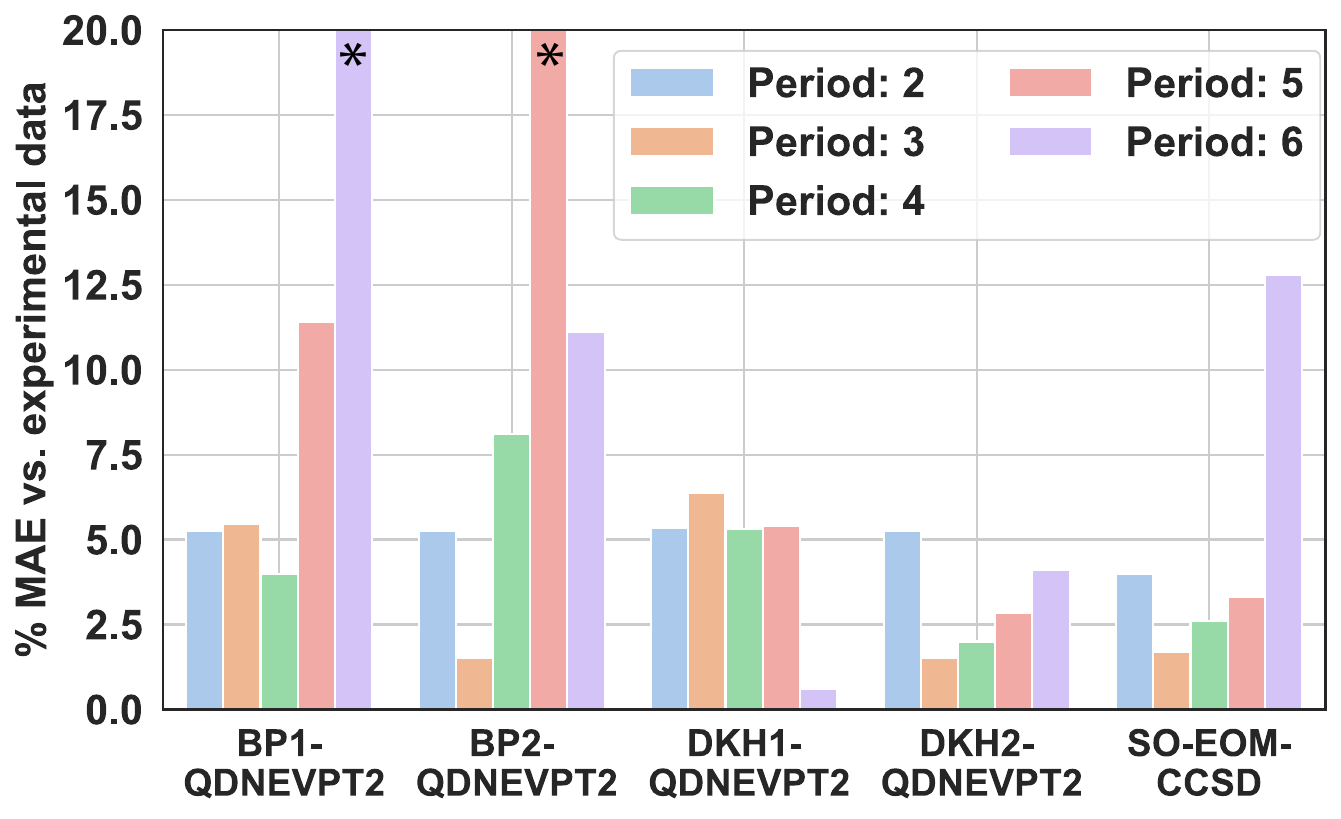}\label{figb:pblock_lan}} 
	\captionsetup{justification=raggedright,singlelinecheck=false}
	\caption{
		Mean absolute errors (MAE, \%) in zero-field splitting for the main group elements and diatomics calculated using the spin--orbit QDNEVPT2 methods and SO-EOM-CCSD\cite{Cheng:2018p044108} relative to the experimental data. 
		MAE are calculated for the chemical systems across each (a) group and (b) period of the periodic table.
		Bars that exceed the scale of the plot are indicated with asterisks.
		See \cref{tab:small_mol} for data on individual systems.
	}
	\label{fig:x2cfig1}
\end{figure*}

\begin{table*}[t]
	\caption{
		Spin--orbit zero-field splitting (\cm) in the ${}^2P$ ground term of atoms and ${}^2\Pi$ ground term of diatomics computed using the spin--orbit QDNEVPT2 methods. 
		Results are compared to the reference data calculated using the EOM-CCSD(SOC) method\cite{Cao:2017p37133721} and experiments.\cite{Moore:1949_book, Radziemski:1969p424443, LucKoenig:1975p199219, Martin:1979p817864, Huber:1982p298, Fink:1989p1928, Ram2000:p915, Kramida:2007p544557,Shirai:2007p509615,Deverall:1953p297299}.
		All methods employed the uncontracted ANO-RCC-VTZP basis set.
	}
	\label{tab:small_mol_2}
	\setstretch{1}
	\footnotesize
	\centering
	\begin{threeparttable}
		\begin{tabular}{cccccccc}
			\hline\hline
			System			& BP1-			& BP2-		& DKH1-		& DKH2-      		& EOM-		& Experiment	\\ 
			& QDNEVPT2		& QDNEVPT2	& QDNEVPT2	& QDNEVPT2	& CCSD(SOC)\cite{Cao:2017p37133721}	&			\\			
			\hline
			\ce{B} 			& 15.0		& 14.5			& 15.0		& 14.5 		& 13.7			& 15.3\cite{Kramida:2007p544557}\\	
			\ce{Al}			& 107.6		& 109.9			& 106.8		& 109.4		& 107.5			& 112\cite{Martin:1979p817864}	\\
			\ce{Ga}			& 887.4		&  867.9			& 840.4		& 818.8		& 797.6			& 826\cite{Shirai:2007p509615}	\\
			\ce{In} 			& 2560.8		& 2859.2 			& 2205.2		& 2219.0		& 2103.6			& 2213\cite{Deverall:1953p297299}	\\
			\ce{Tl} 			& 12475.8		& 8655.5 			& 7745.1 		& 8113.3		& 6794.1			& 7793\cite{Moore:1949_book}	\\\\
			\ce{F} 			& 401.5		& 405.7			& 400.5		& 405.0		& 397.7			& 404\cite{Moore:1949_book}	\\	
			\ce{Cl}			& 789.7		& 867.8			& 779.5		& 858.6		& 876.0			& 882\cite{Radziemski:1969p424443}	\\
			\ce{Br}			& 3574.4		& 3926.0			& 3329.4		& 3625.0		& 3648.8			& 3685\cite{Moore:1949_book}	\\
			\ce{I} 			& 8150.0		& 10343.7 			& 6824.7		& 7581.0		& 7754.6			& 7603\cite{LucKoenig:1975p199219}	\\
			\ce{At} 			& 34153.5		& 345491.0		& 19970.1		& 23002.4		& 24880.5			& --	 	\\\\
			\ce{CH} 			& 29.0		& 27.4			& 29.0		& 27.3		& 27.4			& 27\cite{huber:2013p298}	\\	
			\ce{SiH}			& 128.0		& 136.6			& 127.0		& 135.6		& 139.3			& 142\cite{huber:2013p298}	\\
			\ce{GeH}			& 864.1		& 910.2 			& 815.4		& 854.9		& 882.9			& 892\cite{Huber:1982p298}	\\
			\ce{SnH} 			& 2286.3		& 2713.0			& 1961.6		& 2103.7		& 2187.0			& 2178\cite{Huber:1982p298}	\\\\
			\ce{OH}	    	     	& 152.6    	 	& 123.4            		& 152.3		&  123.2         	& 140.1			& 139\cite{Huber:1982p298}	\\
			\ce{SH}			& 375.6		& 381.7			& 371.4		& 378.2		& 375.3			& 377\cite{Huber:1982p298}	\\
			\ce{SeH}			& 1835.2		& 1931.3			& 1718.1		& 1793.3		& 1742.9			& 1763\cite{Ram2000:p915}	\\
			\ce{TeH}			& 4281.9		& 5212.2			& 3626.1		& 3942.6		& 3913.4			& 3816\cite{Fink:1989p1928} 	\\
			
			\hline\hline
		\end{tabular}
	\end{threeparttable}
\end{table*}

\begin{figure*}[t]
	\subfigure[]{
		\includegraphics[width=0.48\textwidth]{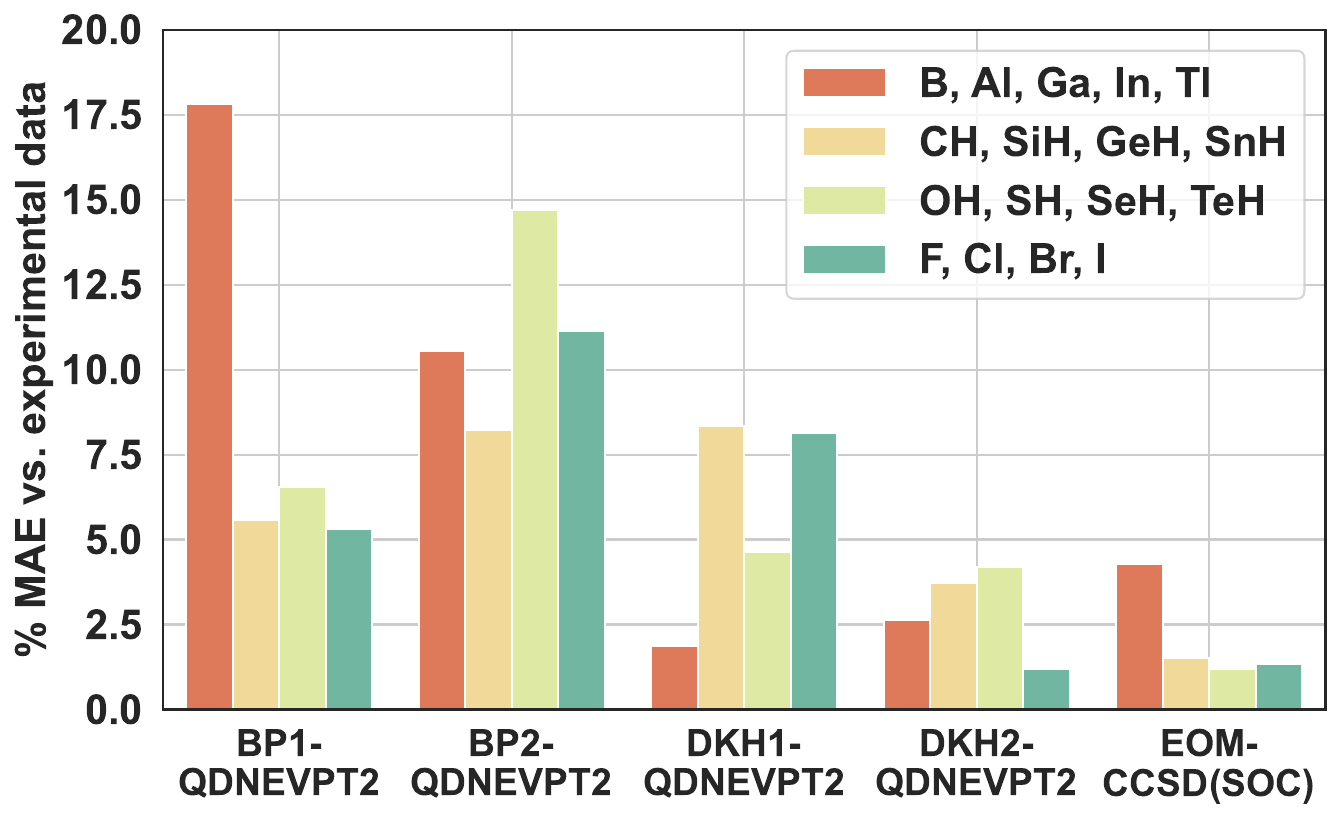}\label{figa:pblock_liu}}  
	\subfigure[]{
		\includegraphics[width=0.48\textwidth]{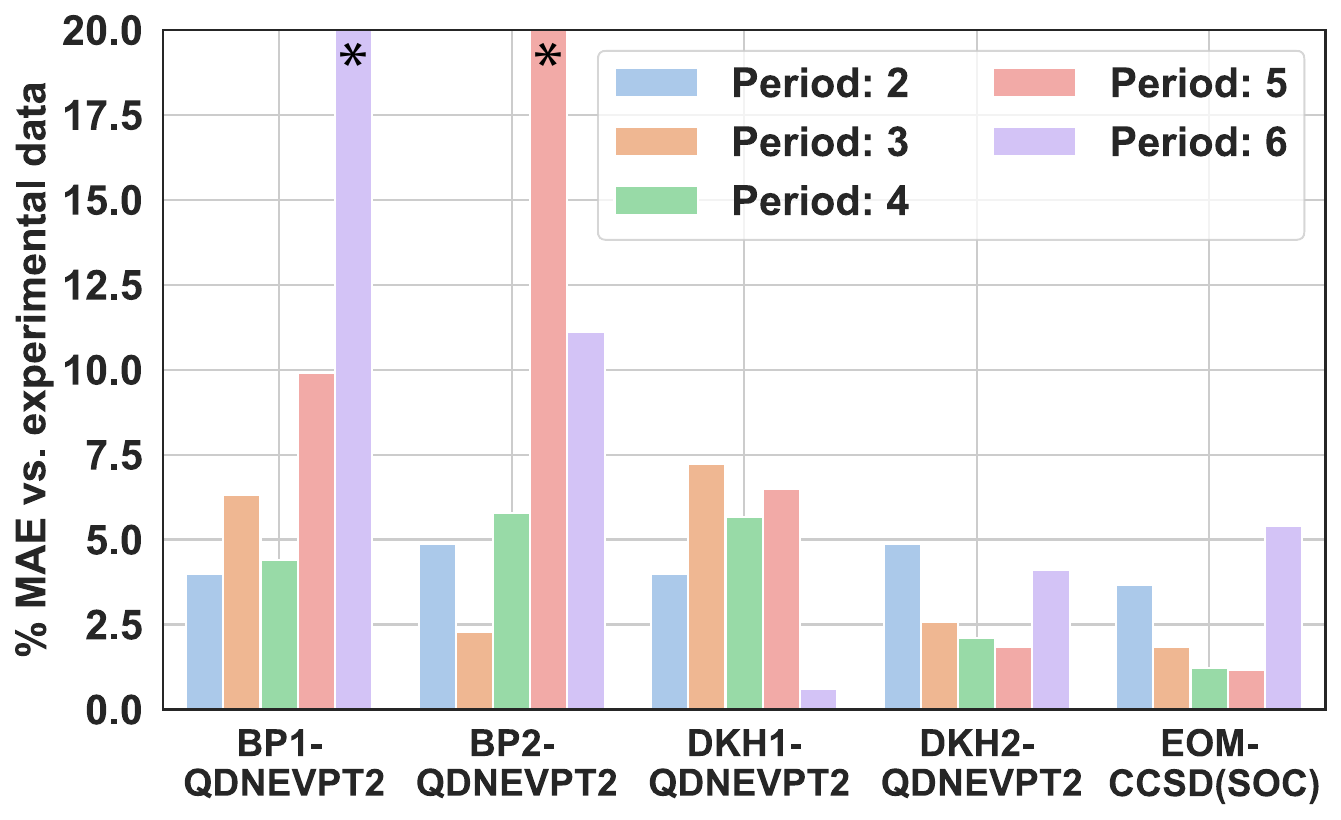}\label{figb:pblock_liu}} 
	\captionsetup{justification=raggedright,singlelinecheck=false}
	\caption{
		Mean absolute errors (MAE, \%) in zero-field splitting for the main group elements and diatomics calculated using the spin--orbit QDNEVPT2 methods and EOM-CCSD(SOC)\cite{Cao:2017p37133721} relative to the experimental data.
		MAE are calculated for the chemical systems across each (a) group and (b) period of the periodic table.
		Bars that exceed the scale of the plot are indicated with asterisks.
		See \cref{tab:small_mol} for data on individual systems.
	}
	\label{fig:x2cfig2}
\end{figure*}

We begin by investigating the accuracy of spin--orbit QDNEVPT2 methods for simulating the zero-field splitting (ZFS) in open-shell atoms and diatomic molecules consisting of main group elements ($p$-block of periodic table), for which accurate theoretical and experimental reference data is available. 
Our first benchmark set consists of 9 atoms and 8 diatomics shown in \cref{tab:small_mol}.
These atoms and molecules possess either the ${}^2P$  or ${}^2\Pi$ ground electronic term, which split into ${}^2P_{1/2}$ and ${}^2P_{3/2}$ or ${}^2\Pi_{1/2}$ and ${}^2\Pi_{3/2}$ energy levels upon incorporating spin--orbit coupling, respectively.
In this benchmark, we employ the uncontracted ANO-RCC basis set and compare the performance of spin--orbit QDNEVPT2 methods to that of spin--orbit equation-of-motion coupled cluster theory with single and double excitations developed by Cheng and co-workers (SO-EOM-CCSD).\cite{Cheng:2018p044108} 
The SO-EOM-CCSD method is a two-component perturbative approach that utilizes the X2C-1e treatment of scalar relativistic effects and mean-field X2C description of spin--orbit coupling, which has a close relationship with the DKH1/DKH2 approach described herein.

The performance of spin--orbit QDNEVPT2 and SO-EOM-CCSD methods in predicting ZFS is compared in \cref{fig:x2cfig1}, where mean absolute errors (MAE, \%) relative to experimental data are computed for atoms and molecules in \cref{tab:small_mol} across each group (a) or period (b) of periodic table.
All four QDNEVPT2 methods show very similar performance for the second period with errors of $\sim$ 5 \%.
Significant differences in computed MAE are observed already for the third period where BP2- and DKH2-QDNEVPT2 show smaller errors ($\sim$ 2 \%) compared to that of BP1- and DKH1-QDNEVPT2 ($\sim$ 5 to 6 \%).
For the fourth period, a large increase in MAE is observed from BP1- to BP2-QDNEVPT2, highlighting the well-known problems of Breit--Pauli Hamiltonian in describing the spin--orbit coupling of elements with heavier nuclei. 
This trend continues for period 5 where BP1- and BP2-QDNEVPT2 exhibit MAE larger than 10 \%.
The DKH-based methods perform reliably for periods 2 to 5, with MAE of $\sim$ 5 \% for DKH1-QDNEVPT2 and  $\lesssim$ 2.5 \% for DKH2-QDNEVPT2, the latter being very close to the MAE of SO-EOM-CCSD.
For the only element from period 6 in this benchmark set (Tl), the best results are shown by DKH1-QDNEVPT2 (0.6 \% error) and DKH2-QDNEVPT2 (4.1 \% error), while SO-EOM-CCSD shows a large error of 12.8 \%.

In \cref{tab:small_mol_2} and \cref{fig:x2cfig2}, we compare the accuracy of spin--orbit QDNEVPT2 methods in calculating ZFS to that of the spin--orbit EOM-CCSD method (EOM-CCSD(SOC)) developed by Cao {\it et al}.\cite{Cao:2017p37133721}
In EOM-CCSD(SOC), the dynamic correlation and spin--orbit coupling effects are incorporated by self-consistently solving the coupled cluster equations utilizing the same two-component Hamiltonian as the one employed in DKH1-QDNEVPT2 (sf-X2C-1e+so-DKH1).
The calculations for this benchmark set were performed using the uncontracted ANO-RCC-VTZP basis to enable direct comparison with the EOM-CCSD(SOC) results. 
Compared to \cref{tab:small_mol}, the data in \cref{tab:small_mol_2} includes ZFS for At (period 6) and group 14 hydrides (OH, SH, SeH, TeH), but does not contain data for the group 17 oxides.

As illustrated in \cref{fig:x2cfig2}, the performance of DKH2-QDNEVPT2 is similar to EOM-CCSD(SOC), which shows somewhat smaller MAE for periods 2 to 5 (by $\sim$ 1 to 1.5 \%), but a larger error for period 6 (by $\sim$ 1 \%).
Meanwhile, DKH1-QDNEVPT2 exhibits significantly larger errors (by a factor of $\sim$ 3) when compared to EOM-CCSD(SOC) for periods 3 to 5, despite using the same two-component Hamiltonian. 
This suggests that the second-order effects in the description of dynamic correlation and spin-orbit coupling incorporated in DKH2-QDNEVPT2 are important to achieve accuracy similar to that of self-consistent two-component relativistic methods such as  EOM-CCSD(SOC).

Overall, our results demonstrate that for the main group elements and their diatomic molecules with predominantly single-reference electronic structure DKH2-QDNEVPT2 shows the highest accuracy for calculating ZFS out of all spin--orbit QDNEVPT2 methods considered in this work.
The DKH1-QDNEVPT2  method exhibits somewhat larger errors in ZFS, but performs reliably for elements across the entire $p$-block of periodic table. 
The BP1- and BP2-QDNEVPT2 implementations start to deteriorate in quality for period 4 and are unreliable for periods 5 and 6.
The accuracy of DKH2-QDNEVPT2 is comparable to that of spin--orbit equation-of-motion coupled cluster methods based on the X2C-type Hamiltonians. 
Although all chemical systems in \cref{tab:small_mol,tab:small_mol_2} have single-reference electronic structure, the QDNEVPT2 methods considered in this work are multireference in nature and are expected to be more reliable than coupled cluster theory for electronic states with strong multiconfigurational character.

\subsection{Transition Metal Elements}
\label{sec:soc_2}

\begin{table*}[t]
	\caption{
		Spin--orbit zero-field splitting (\cm) in the ${}^2D$ ground term of transition metal atoms computed using the spin--orbit QDNEVPT2 methods. 
		Results are compared to the reference data calculated using the X2C-MRCISD method\cite{Hu:2020p29752984} and experiment.\cite{Kaufman:1976p599600}
		Shown in parentheses are the \% errors with respect to experimental results. 
		All methods employed the X2C-TZVPall-2c basis set.
	}
	\label{tab:metal}
	\setstretch{1}
	\footnotesize
	\centering
	\begin{threeparttable}
		\begin{tabular}{ccccccc}
			\hline\hline
			System		& BP1-		& BP2-		& DKH1-		& DKH2-		& X2C-MRCISD\cite{Hu:2020p29752984}			& Experiment\cite{Kaufman:1976p599600}	\\ 
			& QDNEVPT2	& QDNEVPT2	& QDNEPVT2	& QDNEVPT2	& 			&			\\			
			\hline
			\ce{Sc}		& 174.3	(3.6)		& 139.8 (16.9)		& 174.3	(3.6)	& 140.9	(16.3)	& 185.5	(10.2)		& 168.3		\\
			\ce{Y}		& 494.2	(6.8)		& 422.0 (20.4)		& 488.2	(7.9)	& 428.4	(19.2)	& 524.3	(1.1)		& 530.3		\\
			\ce{La}		& 999.9	(5.1)		& 882.9	(16.2)	& 965.6	(8.3)	& 896.6	(14.9)	& 935.6	(11.2)	 		& 1053.2		\\
			\hline\hline
		\end{tabular}
	\end{threeparttable}
\end{table*}

\begin{table*}[t]
	\caption{
		Spin--orbit zero-field splitting (meV) in the ${}^2D$ excited term of \ce{Ag} and \ce{Au} computed using the spin--orbit QDNEVPT2 methods. 
		Results are compared to the data from the X2C-CASSCF and 4C-CASSCF calculations\cite{Sharma:2022p29472954} and experiment.\cite{NIST_ASD}
		Shown in parentheses are the \% errors with respect to experimental results. 
		All methods employed the X2C-TZVPall-2c basis set.
	}
	\label{tab:metal_ii}
	\setstretch{1}
	\footnotesize
	\centering
	\begin{threeparttable}
		\begin{tabular}{ccccccccc}
			\hline\hline
			System			& BP1-		& BP2-		& DKH1-		& DKH2-		& X2C-		& 4C-	 		& Experiment\cite{NIST_ASD}	\\ 
							& QDNEVPT2	& QDNEVPT2	& QDNEVPT2	& QDNEVPT2	& CASSCF\cite{Sharma:2022p29472954}		& CASSCF\cite{Sharma:2022p29472954}		&			\\			
			\hline
			\ce{Ag}		& 542 (2.1)	& 545 (1.6)	& 532 (3.9)	& 540 (2.5)	& 584 (5.4)	& 586 (5.7)	& 554		\\
			\ce{Au}		& 1636 (7.5)	& 1569 (3.1) 	& 1522 (0.1)	& 1519 (0.1)	& 1571 (3.2)	& 1601 (5.2)	& 1521		\\
			\hline\hline
		\end{tabular}
	\end{threeparttable}
\end{table*}

In contrast to the main group elements, most transition metals are known to exhibit significant multireference effects in the ground or excited electronic states.
In \cref{tab:metal,tab:metal_ii}, we apply the spin--orbit QDNEVPT2 methods to the Sc, Y, and La atoms with the ground ${}^2D$  term ($nd^1$ configuration, $n$ = 3, 4, 5) and to the Ag and Au atoms with the excited ${}^2D$ term ($nd^9 (n+1)s^2$ configuration, $n$ = 4, 5).
We compare our results to the available ZFS data from experiments\cite{Kaufman:1976p599600,NIST_ASD} and variational relativistic electronic structure calculations.\cite{Hu:2020p29752984,Sharma:2022p29472954}
All theoretical ZFS were computed using the X2C-TZVPall-2c basis set (see \cref{sec:comp_details} for details).

Incorporating spin--orbit coupling in Sc, Y, and La spits their ground ${}^2D$ term into the ${}^2D_{3/2}$ and ${}^2D_{5/2}$ levels. 
Simulating this ZFS accurately is challenging even for variational electronic structure methods, as demonstrated by the two-component X2C-MRCISD results\cite{Hu:2020p29752984} in \cref{tab:metal} that exhibit large errors relative to the experimental data\cite{Kaufman:1976p599600} (up to 11.2 \%).
Similarly, the ZFS computed using BP2- and DKH2-QDNEVPT2 deviate significantly from the experimental data with errors ranging from 14.9 to 20.4 \%.
Although we cannot quantify the source of these errors, the poor performance of variational X2C-MRCISD method for Sc and La suggests that they are at least in part due to high-order dynamic correlation effects, such as triple (and higher) excitations in non-active orbitals, and their interplay with spin--orbit coupling.
Lowering the level of theory to BP1- and DKH1-QDNEVPT2 fortuitously improves agreement with the experiment, producing errors smaller than those of X2C-MRCISD for Sc and La.

\cref{tab:metal_ii} presents the spin--orbit QDNEVPT2 results for the ZFS in excited ${}^2D$ term of Ag and Au. 
Here, we use the experimental results\cite{NIST_ASD} as the reference and present the data from two- and four-component CASSCF calculations performed by  Sharma {\it et al.}\cite{Sharma:2022p29472954} (X2C-CASSCF and 4C-CASSCF, respectively) that did not incorporate dynamic correlation effects outside the active space.
The highest accuracy is demonstrated by DKH2-QDNEVPT2, which predicts the ${}^2D_{3/2}$ -- ${}^2D_{5/2}$ splitting in Ag and Au with 2.5 \% and 0.1 \% errors, respectively, relative to experiment.
The accuracy of QDNEVPT2 methods decreases in the order DKH2 $>$ DKH1  $>$ BP2  $>$ BP1, with the BP1-QDNEVPT2 errors reaching 7.5\% for Au. 
Except for BP1-QDNEVPT2, all QDNEVPT2 methods agree better with experiment than X2C-CASSCF and 4C-CASSCF, suggesting that including dynamic correlation is quite important for computing accurate ZFS of Ag and Au.

\subsection{Heavy Elements and Molecules}
\label{sec:soc_3}

\begin{table*}[t]
	\caption{
		Spin--orbit zero-field splitting (\cm) in the ${}^2F$ ground term of \ce{U^5+} computed using the spin--orbit QDNEVPT2 methods. 
		Results are compared to the reference data from the X2C-MRCISD calculations\cite{Hu:2020p29752984} and experiment.\cite{Kaufman:1976p599600}
		All methods employed the SARC-DKH2 basis set.
	}
	\label{tab:u}
	\setstretch{1}
	\footnotesize
	\centering
	\begin{threeparttable}
		\begin{tabular}{ccccccc}
			\hline\hline
			System		& BP1-		& BP2-		& DKH1-		& DKH2-		& X2C-		& Experiment\cite{Kaufman:1976p599600}	\\ 
			& QDNEVPT2	& QDNEVPT2	& QDNEPVT2	& QDNEVPT2	& MRCISD\cite{Hu:2020p29752984}				&			\\			
			\hline
			\ce{U(V)}		& 8170.8			& 7144.1		& 8038.2		& 7316.4		& 7863.9			& 7605.8		\\
			\hline\hline
		\end{tabular}
	\end{threeparttable}
\end{table*}

\begin{table*}[t]
	\caption{
		Excited-state energies (\cm) of \ce{NpO2^2+} computed using the spin--orbit QDNEVPT2 methods.
		Results are compared to the reference data from the SO-SHCI calculations.\cite{Wang:2023p848855}
		For all methods, the uncontracted ANO-RCC-VTZP and cc-pVTZ basis sets were used for the \ce{Np} and \ce{O} atoms, respectively.
	} 
	\label{tab:npo2}
	\setstretch{1}
	\footnotesize
	\centering
	\begin{threeparttable}
		\begin{tabular}{c c c c c c}
			\hline\hline
			Electronic   		& BP1- 		& BP2- 		& DKH1-		& DKH2-		& SO-SHCI\cite{Wang:2023p848855} 	\\ 
			state & QDNEVPT2	& QDNEVPT2	&QDNEPVT2	& QDNEVPT2	&		     \\
			\hline
			${}^2\Phi_{5/2u}$  		& 0.0         	& {0.0	}           	& 0.0                 & 0.0			& 0.0              \\
			${}^2\Delta_{3/2u}$   	& 3603.7		& {3025.8}	  	& 3570.5	        & 3595.1		& 3429            \\
			${}^2\Phi_{7/2u}$  		& 8057.2		& {3162.6}	   	& 7916.3            & 7608.6		& 7165            \\
			${}^2\Delta_{5/2u}$   	& 9238.4		& {3288.4}	      	& 9100.7            & 8956.7		& 8868            \\ \hline\hline
		\end{tabular}			
	\end{threeparttable}
\end{table*} 

\begin{table*}[t]
	\caption{
		Excited-state energies (\cm) of  \ce{UO2^+} computed using the spin--orbit QDNEVPT2 methods.
		Results are compared to the data from CASPT2-SO calculations\cite{Gendron:2014p8577} and experiment.\cite{Merritt:2008p84301}
		The contracted ANO-RCC-VTZP basis set was employed in all calculations. 
	} 
	\label{tab:uo2}
	\setstretch{1}
	\footnotesize
	\centering
	\begin{threeparttable}
		\begin{tabular}{c c c c c c c}
			\hline\hline
			Electronic    	   		& BP1- 		& BP2- 		& DKH1-		& DKH2-		& CASPT2-	& Experiment\cite{Merritt:2008p84301} 	\\ 
			state					& QDNEVPT2	& QDNEVPT2	& QDNEPVT2	& QDNEVPT2	& SO\cite{Gendron:2014p8577} 	& 	\\
			\hline
			${}^2\Phi_{5/2u}$  		& 0.0         	& 0.0	           	& 0.0                 & 0.0			& 0.0              	& 0			\\
			${}^2\Delta_{3/2u}$   	& 2912.1	        	& 2838.2	  	& 2922.9		& 2862.0		& 2616             	& 2658		\\
			${}^2\Phi_{7/2u}$  		& 6471.7	        	& 6187.3	   	& 6429.7            & 6136.4		& 6679             	& \textendash	  \\
			${}^2\Delta_{5/2u}$   	& 7905.2	       	& 7668.8	      	& 7918.8            & 7653.1		& 7889            	& \textendash 	 \\
			\hline\hline
		\end{tabular}			
	\end{threeparttable}
\end{table*} 

Finally,  we consider \ce{U^5+}, \ce{NpO2^2+}, and \ce{UO2^+}, which contain actinide elements that are challenging for perturbative two-component relativistic theories due to strong spin--orbit coupling and nearly degenerate partially filled $f$-orbitals in their electronic states.\cite{Ruiperez:2009p14201428,Gendron:2014p79948011,Knecht:2016p58815894,Mussard:2018p154,Hu:2020p29752984}

\cref{tab:u} presents the spin--orbit QDNEVPT2 results for the ${}^2F$ ground term of \ce{U^5+} originating from the $5f^1$ electronic configuration.
As a reference, we employ the experimental ZFS reported by Kaufman {\it et al}.\cite{Kaufman:1976p599600} and the theoretical data from variational X2C-MRCISD calculations by Hu {\it et al}.\cite{Hu:2020p29752984}
For this system, all computations were performed using the SARC-DKH2 basis set.
DKH2-QDNEVPT2 shows the best agreement with experiment out of all perturbative methods, underestimating the experimental ZFS by 3.8 \%, which is similar to the error of X2C-MRCISD (3.4 \%).
As for Ag and Au, the accuracy of spin--orbit QDNEVPT2 methods decreases in the order DKH2 (3.8 \% error) $>$ DKH1 (5.7 \%) $>$ BP2 (6.1 \%)  $>$ BP1 (7.4 \%), demonstrating that the second-order description of dynamical correlation and spin--orbit coupling using the DKH2 Hamiltonian is essential for achieving accuracy similar to X2C-MRCISD.

Next, we use spin--orbit QDNEVPT2 to compute the energies of excited states originating from the zero-field splitting in the ${}^2\Phi$ and ${}^2\Delta$ terms of \ce{NpO2^2+}, which exhibit strong electron correlation and spin--orbit coupling effects (\cref{tab:npo2}). 
In this study, we benchmark against the recently published results of SO-SHCI calculations\cite{Wang:2023p848855} that utilized the variational two-component treatment of relativistic effects with the DKH1 Hamiltonian.
We note that the SO-SHCI calculations were performed in the (13e, 60o) active space, while our spin--orbit QDNEVPT2 methods correlated all 107 electrons in 433 molecular orbitals thus providing a more accurate description of dynamic correlation.
Using the same basis set and molecular geometry as in the SO-SHCI study, the best agreement with the reference data is achieved by the DKH2-QDNEVPT2 method with the largest error of 6.2 \% (444 \cm) for ${}^2\Phi_{7/2u}$.
The error in ${}^2\Phi_{7/2u}$ excitation energy increases when using DKH1-QDNEVPT2 (10.4 \%) or BP1-QDNEVPT2 (12.5 \%).
The BP2-QDNEVPT2 method yields severely underestimated excitation energies despite using the same reference SA-CASSCF wavefunction as the other spin--orbit QDNEVPT2 calculations.

In \cref{tab:uo2}, we also report the excited-state energies for \ce{UO2^+}, which has the same electronic states and configuration as \ce{NpO2^2+}.
We compare the spin--orbit QDNEVPT2 results to the data from experimental measurements for the ${}^2\Delta_{3/2u}$ state\cite{Merritt:2008p84301} and perturbative CASPT2-SO calculations\cite{Gendron:2014p8577} utilizing the same basis set and structural parameters.
Interestingly, we find that for this system BP2- and DKH2-QDNEVPT2 yield similar results, in a closer agreement to the  experimental ${}^2\Delta_{3/2u}$ energy than BP1- and DKH1-QDNEVPT2, despite BP2-QDNEVPT2 showing large errors for  \ce{NpO2^2+}.
This uneven performance of BP2-QDNEVPT2 is likely associated with the low-$Z$ nature of approximations in the Breit--Pauli Hamiltonian and warrants further investigation.

\section{Conclusions}
\label{sec:conclusions}

In this work, we developed a formulation of quasidegenerate N-electron valence perturbation theory (QDNEVPT) that enables consistent second-order treatment of dynamic correlation and spin-orbit coupling for chemical systems with multiconfigurational electronic structure.
Utilizing the Breit--Pauli (BP) and exact two-component Douglas--Kroll--Hess (DKH) relativistic Hamiltonians, the resulting approaches termed BP2- and DKH2-QDNEVPT2 have computational cost similar to that of conventional non-relativistic QDNEVPT2.
Although derived from perturbation theory, the BP2- and DKH2-QDNEVPT2 methods compute the energies and wavefunctions of electronic states by diagonalizing an effective Hamiltonian, which delivers the exact eigenvalues and eigenstates of BP and DKH2 Hamiltonians in the limit of full configuration interaction.
By expanding the treatment of dynamic correlation and spin-dependent relativistic effects to second order, BP2- and DKH2-QDNEVPT2 allow to obtain the accurate energies and wavefunctions of spin--orbit-coupled states with compact non-relativistic representations of effective Hamiltonian.
To quantify the importance of second-order effects, we also considered QDNEVPT2 with the first-order BP and DKH treatment of spin--orbit coupling, denoted as BP1- and DKH1-QDNEVPT2, respectively.

Our results demonstrate that, out of four spin--orbit QDNEVPT2 approaches studied in this work, DKH2-QDNEVPT2 provides the most accurate and reliable description of zero-field splitting for a variety of chemical systems, including main group elements, transition metal atoms, actinides, and their compounds. 
For the main group elements with single-reference electronic structures, the accuracy of DKH2-QDNEVPT2 is similar to that of two-component equation-of-motion coupled cluster theory with single and double excitations. 
When applied to the Ag and Au transition metal atoms, DKH2-QDNEVPT2 shows higher accuracy than exact two-component (X2C-) complete active space self-consistent field method, but exhibits larger errors than the X2C implementation of multireference configuration interaction with singles and doubles (X2C-MRCISD) for Sc, Y, and La. 
For heavier elements and their compounds (\ce{U^5+}, \ce{NpO2^2+}, and \ce{UO2^+}),  DKH2-QDNEVPT2 delivers results of the quality similar to  that of  X2C-MRCISD and spin--orbit implementation of semistochastic heat-bath CI (SO-SHCI).
The DKH1-QDNEVPT2 method tends to show larger errors than DKH2-QDNEVPT2 by $\sim$ 2 to 3 \% relative to experimental results.
The BP1- and BP2-QDNEVPT2 implementations exhibit accurate performance for the second- and third-period elements, but become increasingly inaccurate and unreliable for heavier atoms and molecules.

Overall, the DKH2-QDNEVPT2 method developed in this work shows promise as an accurate electronic structure approach that incorporates multireference effects, dynamic correlation, and spin--orbit coupling with affordable computational cost.
Applications of DKH2-QDNEVPT2 to chemical systems larger than the ones presented in this study necessitate its efficient computer implementation. 
Other developments of this approach can be envisioned, such as extensions to simulate spin-dependent and magnetic properties, high-energy states, and nonradiative decay rates.

\suppinfo
Structural parameters, active spaces, and the states included in SA-CASSCF calculations. 

\acknowledgement
This work was supported by the National Science Foundation under Grant No.\@ CHE-2044648.
Computations were performed at the Ohio Supercomputer Center under Project No.\@ PAS1583.\cite{OhioSupercomputerCenter1987}
The authors would like to thank Professor Lan Cheng for insightful discussions.


\providecommand{\latin}[1]{#1}
\makeatletter
\providecommand{\doi}
  {\begingroup\let\do\@makeother\dospecials
  \catcode`\{=1 \catcode`\}=2 \doi@aux}
\providecommand{\doi@aux}[1]{\endgroup\texttt{#1}}
\makeatother
\providecommand*\mcitethebibliography{\thebibliography}
\csname @ifundefined\endcsname{endmcitethebibliography}
  {\let\endmcitethebibliography\endthebibliography}{}

\end{document}